\documentclass[aps,prd,superscriptaddress,showpacs,floatfix,twocolumn]{revtex4}

\usepackage{graphicx}

\begin{document}

\title{Mid-Rapidity Direct-Photon Production in p+p Collisions at $\sqrt{s}=200$~GeV}

\date{\today}

\newcommand{\abilene}{Abilene Christian University, Abilene, TX 79699, USA}
\newcommand{\acadsin}{Institute of Physics, Academia Sinica, Taipei 11529, Taiwan}
\newcommand{\banaras}{Department of Physics, Banaras Hindu University, Varanasi 221005, India}
\newcommand{\barc}{Bhabha Atomic Research Centre, Bombay 400 085, India}
\newcommand{\bnl}{Brookhaven National Laboratory, Upton, NY 11973-5000, USA}
\newcommand{\caucr}{University of California - Riverside, Riverside, CA 92521, USA}
\newcommand{\ciae}{China Institute of Atomic Energy (CIAE), Beijing, People's Republic of China}
\newcommand{\cns}{Center for Nuclear Study, Graduate School of Science, University of Tokyo, 7-3-1 Hongo, Bunkyo, Tokyo 113-0033, Japan}
\newcommand{\columbia}{Columbia University, New York, NY 10027 and Nevis Laboratories, Irvington, NY 10533, USA}
\newcommand{\dapnia}{Dapnia, CEA Saclay, F-91191, Gif-sur-Yvette, France}
\newcommand{\debrecen}{Debrecen University, H-4010 Debrecen, Egyetem t{\'e}r 1, Hungary}
\newcommand{\fsu}{Florida State University, Tallahassee, FL 32306, USA}
\newcommand{\gsu}{Georgia State University, Atlanta, GA 30303, USA}
\newcommand{\hiroshima}{Hiroshima University, Kagamiyama, Higashi-Hiroshima 739-8526, Japan}
\newcommand{\ihepprot}{Institute for High Energy Physics (IHEP), Protvino, Russia}
\newcommand{\isu}{Iowa State University, Ames, IA 50011, USA}
\newcommand{\jinrdubna}{Joint Institute for Nuclear Research, 141980 Dubna, Moscow Region, Russia}
\newcommand{\kaeri}{KAERI, Cyclotron Application Laboratory, Seoul, Korea}
\newcommand{\kangnung}{Kangnung National University, Kangnung 210-702, Korea}
\newcommand{\kek}{KEK, High Energy Accelerator Research Organization, Tsukuba-shi, Ibaraki-ken 305-0801, Japan}
\newcommand{\kfki}{KFKI Research Institute for Particle and Nuclear Physics (RMKI), H-1525 Budapest 114, POBox 49, Hungary}
\newcommand{\korea}{Korea University, Seoul, 136-701, Korea}
\newcommand{\kurchatov}{Russian Research Center ``Kurchatov Institute", Moscow, Russia}
\newcommand{\kyoto}{Kyoto University, Kyoto 606-8502, Japan}
\newcommand{\labllr}{Laboratoire Leprince-Ringuet, Ecole Polytechnique, CNRS-IN2P3, Route de Saclay, F-91128, Palaiseau, France}
\newcommand{\lawllnl}{Lawrence Livermore National Laboratory, Livermore, CA 94550, USA}
\newcommand{\losalamos}{Los Alamos National Laboratory, Los Alamos, NM 87545, USA}
\newcommand{\lpc}{LPC, Universit{\'e} Blaise Pascal, CNRS-IN2P3, Clermont-Fd, 63177 Aubiere Cedex, France}
\newcommand{\lund}{Department of Physics, Lund University, Box 118, SE-221 00 Lund, Sweden}
\newcommand{\muenster}{Institut f\"ur Kernphysik, University of Muenster, D-48149 Muenster, Germany}
\newcommand{\myongji}{Myongji University, Yongin, Kyonggido 449-728, Korea}
\newcommand{\nagasaki}{Nagasaki Institute of Applied Science, Nagasaki-shi, Nagasaki 851-0193, Japan}
\newcommand{\newmex}{University of New Mexico, Albuquerque, NM 87131, USA}
\newcommand{\nmsu}{New Mexico State University, Las Cruces, NM 88003, USA}
\newcommand{\ornl}{Oak Ridge National Laboratory, Oak Ridge, TN 37831, USA}
\newcommand{\orsay}{IPN-Orsay, Universite Paris Sud, CNRS-IN2P3, BP1, F-91406, Orsay, France}
\newcommand{\pnpi}{PNPI, Petersburg Nuclear Physics Institute, Gatchina, Russia}
\newcommand{\riken}{RIKEN (The Institute of Physical and Chemical Research), Wako, Saitama 351-0198, JAPAN}
\newcommand{\rikjrbrc}{RIKEN BNL Research Center, Brookhaven National Laboratory, Upton, NY 11973-5000, USA}
\newcommand{\saispbstu}{St. Petersburg State Technical University, St. Petersburg, Russia}
\newcommand{\saopaulo}{Universidade de S{\~a}o Paulo, Instituto de F\'{\i}sica, Caixa Postal 66318, S{\~a}o Paulo CEP05315-970, Brazil}
\newcommand{\seoulnat}{System Electronics Laboratory, Seoul National University, Seoul, Korea}
\newcommand{\stonybrkc}{Chemistry Department, Stony Brook University, SUNY, Stony Brook, NY 11794-3400, USA}
\newcommand{\stonycrkp}{Department of Physics and Astronomy, Stony Brook University, SUNY, Stony Brook, NY 11794, USA}
\newcommand{\subatech}{SUBATECH (Ecole des Mines de Nantes, CNRS-IN2P3, Universit{\'e} de Nantes) BP 20722 - 44307, Nantes, France}
\newcommand{\tenn}{University of Tennessee, Knoxville, TN 37996, USA}
\newcommand{\titech}{Department of Physics, Tokyo Institute of Technology, Tokyo, 152-8551, Japan}
\newcommand{\tsukuba}{Institute of Physics, University of Tsukuba, Tsukuba, Ibaraki 305, Japan}
\newcommand{\vandy}{Vanderbilt University, Nashville, TN 37235, USA}
\newcommand{\waseda}{Waseda University, Advanced Research Institute for Science and Engineering, 17 Kikui-cho, Shinjuku-ku, Tokyo 162-0044, Japan}
\newcommand{\weizmann}{Weizmann Institute, Rehovot 76100, Israel}
\newcommand{\yonsei}{Yonsei University, IPAP, Seoul 120-749, Korea}
\affiliation{\abilene}
\affiliation{\acadsin}
\affiliation{\banaras}
\affiliation{\barc}
\affiliation{\bnl}
\affiliation{\caucr}
\affiliation{\ciae}
\affiliation{\cns}
\affiliation{\columbia}
\affiliation{\dapnia}
\affiliation{\debrecen}
\affiliation{\fsu}
\affiliation{\gsu}
\affiliation{\hiroshima}
\affiliation{\ihepprot}
\affiliation{\isu}
\affiliation{\jinrdubna}
\affiliation{\kaeri}
\affiliation{\kangnung}
\affiliation{\kek}
\affiliation{\kfki}
\affiliation{\korea}
\affiliation{\kurchatov}
\affiliation{\kyoto}
\affiliation{\labllr}
\affiliation{\lawllnl}
\affiliation{\losalamos}
\affiliation{\lpc}
\affiliation{\lund}
\affiliation{\muenster}
\affiliation{\myongji}
\affiliation{\nagasaki}
\affiliation{\newmex}
\affiliation{\nmsu}
\affiliation{\ornl}
\affiliation{\orsay}
\affiliation{\pnpi}
\affiliation{\riken}
\affiliation{\rikjrbrc}
\affiliation{\saispbstu}
\affiliation{\saopaulo}
\affiliation{\seoulnat}
\affiliation{\stonybrkc}
\affiliation{\stonycrkp}
\affiliation{\subatech}
\affiliation{\tenn}
\affiliation{\titech}
\affiliation{\tsukuba}
\affiliation{\vandy}
\affiliation{\waseda}
\affiliation{\weizmann}
\affiliation{\yonsei}
\author{S.S.~Adler}	\affiliation{\bnl}
\author{S.~Afanasiev}	\affiliation{\jinrdubna}
\author{C.~Aidala}	\affiliation{\bnl}
\author{N.N.~Ajitanand}	\affiliation{\stonybrkc}
\author{Y.~Akiba}	\affiliation{\kek} \affiliation{\riken}
\author{J.~Alexander}	\affiliation{\stonybrkc}
\author{R.~Amirikas}	\affiliation{\fsu}
\author{L.~Aphecetche}	\affiliation{\subatech}
\author{S.H.~Aronson}	\affiliation{\bnl}
\author{R.~Averbeck}	\affiliation{\stonycrkp}
% ...
\collaboration{PHENIX Collaboration} \noaffiliation

%

\begin{abstract}
  A measurement of direct photons in p+p collisions at $\sqrt{s}=
  200$~GeV is presented. A photon excess above background from $\pi^0
  \rightarrow \gamma+\gamma$, $\eta\rightarrow \gamma+\gamma$ and
  other decays is observed in the transverse momentum range $5.5
  <p_\mathrm{T} < 7$~GeV/$c$. The result is compared to a
  next-to-leading-order perturbative QCD calculation.  Within errors,
  good agreement is found between the QCD calculation and the measured
  result.
\end{abstract}
\maketitle

Measurements of particle production at large transverse momenta
($p_\mathrm{T}$) in hadronic interactions provide the possibility to
test perturbative quantum chromodynamics (pQCD). Neutral-pion
production in p+p collisions at $\sqrt{s}=200$~GeV in the range $2 <
p_\mathrm{T} < 13$~GeV/$c$ measured by the PHENIX experiment at RHIC
can be well described by next-to-leading-order (NLO) pQCD
\cite{run2_pp_pi0}.  This comparison, however, relies on the choice of
the parton-to-pion fragmentation function. Measurement of
direct-photon production provides a more direct test of pQCD.
Quark-antiquark annihilation ($q\bar{q} \rightarrow \gamma g$) and
quark-gluon Compton scattering ($qg\rightarrow q \gamma$) contribute
to direct-photon production at leading order \cite{Fritzsch:1977eq}.
Due to the latter process, which dominates the production, the
measurement of direct photons can be used to obtain information on the
parton distribution function of the gluon inside the proton.

Previous direct-photon measurements in p+p collisions were made up to
energies of $\sqrt{s}=63$~GeV (see {\em e.g.}
\cite{Anassontzis:1982gm, Apanasevich:2004dr, Ballocchi:1998au}). For
$\mathrm{p}+\bar{\mathrm{p}}$ collisions direct-photon data are
available at considerably higher energies, $\sqrt{s}$ = 546, 630
GeV~\cite{Albajar:1988im,Ansari:1988te,Alitti:1991yk} up to
$\sqrt{s}=1800$~GeV \cite{Abe:1994rr, Abbott:1999kd}.  At these
energies NLO pQCD calculations describe the direct-photon data within
about 20\%, although systematic differences in the spectral shapes
were observed \cite{Acosta:2002ya}. At energies below
$\sqrt{s}=63$~GeV the agreement between NLO pQCD and data is generally
worse.  With phenomenological approaches based on soft-gluon radiation
of the incoming parton, which leads to an additional transverse
momentum $k_T$, a better description of the data can be obtained
\cite{Apanasevich:1998ki}.  The Relativistic Heavy Ion Collider (RHIC)
at the Brookhaven National Laboratory provides p+p collisions at
energies between the existing data sets, allowing better constraints
on the processes affecting incoming partons.  A further incentive to
study direct-photon production in p+p at $\sqrt{s}=200$~GeV comes from
the measurement of direct photons in collisions of gold nuclei at the
same center-of-mass energy per nucleon-nucleon pair
\cite{Frantz:2004gg, ppg042:2005}.  In central Au+Au collisions
high-$p_\mathrm{T}$ neutral-pion production is suppressed
\cite{Adler:2003qi} which is due to energy loss of the scattered
quarks and gluons in the hot and dense fireball created in these
collisions (jet quenching) \cite{Gyulassy:2003mc}. As direct photons
are not subject to the strong interaction they should not be
suppressed in the jet-quenching model. In this context, the p+p
direct-photon results serve as a baseline against which possible
nuclear effects can be identified.

The data presented in this report were collected during the 2001-2002
run period (Run 2) of RHIC. The neutral-pion cross sections obtained
from this data set were published in \cite{run2_pp_pi0}. The
unpolarized neutral-pion spectrum and the unpolarized direct-photon
spectrum presented here were obtained by averaging over proton bunches
with varying vertical polarization delivered by RHIC.

Direct photons and background photons from decays $\pi^0\rightarrow
\gamma+\gamma$ and $\eta \rightarrow \gamma+\gamma$ were measured with
the electromagnetic calorimeter (EMCal) of the PHENIX experiment
\cite{nim_phenix}. The background from charged particles was
subtracted with the aid of a layer of multi-wire proportional chambers
with pad readout (PC3) which was located directly in front of the
EMCal.  The minimum-bias trigger was provided by two beam-beam
counters (BBC) which were also used to determine the collision vertex.
In addition to the minimum-bias trigger conditions, a
high-$p_\mathrm{T}$ photon trigger was used, derived from the analog
energy signal measured with the EMCal.

The two BBC's were located symmetrically around the nominal
interaction point at $\pm\:1.44$~m along the beamline. The BBC's
subtended the pseudorapidity range $\pm\:(3.1-3.9)$ with full
azimuthal coverage. The collision vertex was determined by measuring
the difference of particle arrival times in the two BBC's.  This
analysis was restricted to events with a vertex in the range
$\pm\:30$~cm. The BBC's were calibrated as a luminosity detector with
absolute luminosity measurements based on the van der Meer scan
technique \cite{Meer}. With these scans the cross section for firing
the BBC minimum-bias trigger was determined to be $21.8 \pm 2.1$~mb
\cite{run2_pp_pi0}. Thus, roughly 50\% of the inelastic p+p events
satisfy the minimum-bias trigger condition if an inelastic p+p cross
section of 42~mb at $\sqrt{s}=200$~GeV is assumed.

The PHENIX EMCal comprises two arms each with 4 sectors
\cite{nim_emc}. The EMCal consists of two different sub-detectors, a
lead-scintillator calorimeter (PbSc, 6 sectors) and a lead-glass
calorimeter (PbGl, 2 sectors). Each sector covers a pseudorapidity
range of $|\eta| < 0.35$ and an azimuthal range of $\Delta\phi \approx
22.5^{\circ}$. Each PbSc (PbGl) sector is highly segmented and
consists of $72 \times 36$ ($96 \times 48$) individual detector
modules, called towers, with a lateral size of $5.5 \times 5.5$~cm$^2$
($4 \times 4$~cm$^2$). With a radial distance of the sectors to the
beamline of roughly 5~m this corresponds to a segmentation of $\Delta
\phi \times \Delta \eta \approx 0.01 \times 0.01$ such that the two
decay photons of a $\pi^0$ are well separated up to neutral-pion
momenta of $p_\mathrm{T} \approx 20$~GeV/$c$.  The different detection
mechanisms of the two sub-detectors (measurement of scintillation
light in PbSc and detection of Cherenkov photons in PbGl) result in a
different response to hadrons.  Thus, the PbSc and PbGl provide photon
measurements with different systematic uncertainties. The energy
calibration of the detector was obtained from the position of the
$\pi^0$ invariant-mass peaks. A $\sim4$\% ($\sim5$\%) shift of the
$\pi^0$ peak position due to energy smearing in conjunction with the
influence of the steeply falling $\pi^0$ $p_\mathrm{T}$ spectrum was
taken into account in the PbSc (PbGl) calibration. The calibration was
corroborated by correlating the EMCal energy with the momentum of
electrons measured with the PHENIX tracking detectors and, in case of
the PbSc, by measuring the energy deposited by minimum-ionizing
particles. From these studies the systematic uncertainty of the energy
measurement was estimated to be less than 1.5\%.  In a direct-photon
analysis it is essential to exclude bad detector modules ("hot
towers") which might give rise to spurious direct-photon signals. A
detailed quality assessment was carried out to identify such towers.
 
The EMCal high-$p_\mathrm{T}$ trigger (called $2 \times 2$) was based
on the analog energy signal measured in $2 \times 2$ groups of
adjacent EMCal towers (called trigger tiles). The average threshold of
the trigger corresponded to an energy signal of 0.75~GeV.  The
probability as a function of the photon $p_\mathrm{T}$ to fire the
trigger was determined by Monte Carlo simulations which included the
variation of the trigger tile thresholds, the EMCal detector response,
and the geometry of the active trigger tiles. This trigger efficiency
was confirmed with minimum-bias data. The photon trigger efficiencies
for PbSc and PbGl reached a plateau above $p_\mathrm{T} =
1.5-2$~GeV/$c$ at the limit of about 0.78 expected from the number of
active towers and $2 \times 2$ trigger tiles. The
high-$p_\mathrm{T}$-trigger photon sample was used above $p_\mathrm{T}
= 3$~GeV/$c$ in the final spectrum.

Another EMCal trigger which did not require a coincidence with the
minimum-bias trigger was used to account for the bias on the particle
measurement due to the minimum-bias event selection.  To this end the
fraction of $\pi^0$'s measured with this EMCal trigger for events
which in addition satisfied the minimum-bias trigger condition was
determined to be $f = 0.75 \pm 0.02$. The unbiased photon and neutral
pions cross sections were then determined by dividing the total number
of measured photons and neutral pions by this number.

The minimum-bias data sample in this analysis consisted of 
$16.7$ million events, corresponding to an integrated luminosity 
of $0.77$~nb$^{-1}$. About 1 in 47 minimum-bias events also satisfied 
the $2 \times 2$ high-$p_\mathrm{T}$ trigger condition. The $18.7$ million 
analyzed $2 \times 2$ events thus corresponded to an integrated 
luminosity of 40.3~nb$^{-1}$. 

The first step in the direct-photon analysis was to define a sample of
direct-photon-candidate hits. An EMCal hit was rejected as a
direct-photon candidate if it formed an invariant mass in the $\pi^0$
or $\eta$ range with other hits in the same or adjacent sectors.  The
invariant-mass window was $110 < m_{\gamma\gamma} < 170$~MeV/$c^2$ for
the $\pi^0$ and $500 < m_{\gamma\gamma} < 620$~MeV/$c^2$ for the
$\eta$, corresponding roughly to a $\pm 2\sigma$ window around the
observed $\pi^0$ and $\eta$ peaks. To keep the rate of accidental
rejections of genuine direct photons low it was required that the
partner hits had a transverse momentum of $p_\mathrm{T} >
0.4$~GeV/$c$. This cut effectively corresponded to a
$p_\mathrm{T}$-dependent upper limit on the energy asymmetry $\alpha =
|E_{\gamma1} - E_{\gamma2}|/(E_{\gamma1} + E_{\gamma2})$ in the
rejection procedure. In spite of this requirement, a small fraction of
genuine direct photons is rejected. This was studied by inserting
artificially generated direct-photon hits into real events.  In order
to keep the hit multiplicity constant, a randomly selected real hit
was removed from an event in this procedure. It was found that the
loss of genuine direct photons was less than 2\% for $p_\mathrm{T} >
3$~GeV/$c$. The final direct-photon spectrum was corrected for this
effect.  In order to increase the chances of finding the partner
photon for a $\pi^0$ or $\eta$ decay photon, direct-photon candidates
were required to lie within a restricted fiducial area which was
defined by a minimum distance of 16 (20) towers to the edge of the
detector for the PbSc (PbGl). For example, for $\pi^0$'s with
$p_\mathrm{T} = 4$~GeV/$c$ and the requirement that one decay photon
has a $p_\mathrm{T} > 0.4$~GeV/$c$ the average distance of the decay
photons in tower units is $\sim 8$ ($\sim 11$) for PbSc (PbGl). With
the chosen fiducial area basically all decay photons from neutral
pions with $p_\mathrm{T} \gtrsim 4$~GeV/$c$ can be tagged.  Monte
Carlo studies showed that the rejection of direct-photon candidates
based on the $\pi^0$ and $\eta$ tagging lead to a reduction of
background photons from hadron decays in the fiducial area of about a
factor of 2 for $p_\mathrm{T} > 5 $~GeV/$c$.  Some direct-photon
analyses only measure isolated direct photons for which the total
transverse energy or the number of charged tracks in a cone centered
around the direct photon is required to lie below a threshold.  No
such cut was used in this analysis.

In order to reduce the background from hadronic hits in the EMCal,
cuts were applied on the lateral shower shape and on the
time-of-flight of the hits. The remaining contamination of charged
particles was subtracted on a statistical basis by employing the PC3
as a charged-particle veto detector. The intrinsic efficiency of the
PC3 for detecting a charged particle was higher than 99\% and the
active PC3 area in the EMCal acceptance was roughly $90\%$. PC3 hits
were projected onto the EMCal surface using a straight line given by
the PC3 hit and the event vertex. An EMCal hit within a certain veto
radius was counted as a charged hit.  The chosen veto radius decreased
with increasing $p_\mathrm{T}$ and for $p_\mathrm{T} > 0.8$~GeV/$c$ a
constant value of 15~cm was used.  The fraction of charged hits was
corrected for random associations with the help of a mixed-event
technique.  The charged-particle background in the
direct-photon-candidate sample was $\sim15$\% around $p_\mathrm{T} =
5$~GeV/$c$ for both PbSc and PbGl. A large fraction of these
background hits, however, comes from photon conversion in the
field-free region between the vertex and PC3. The photon loss due to
conversion was calculated based on the material budget up to PC3.  The
photon conversion probability was 4.1\% for the 2 PbSc sectors in the
East Arm of the central spectrometer, 5.3\% for the 4 sectors in the
West Arm, and 7.4\% for the PbGl. These conversion losses were taken
into account in the final photon cross section.  The correction for
the contamination of the raw spectrum of neutral EMCal hits with
neutrons and anti-neutrons was determined with a detailed GEANT
simulation \cite{geant}. In the case of the PbGl calorimeter the
simulation was based on the creation of Cherenkov photons in order to
achieve a realistic description of the detector response.  The
background from neutral particles was found to decrease with
$p_\mathrm{T}$ and was already less than 
1\% for $p_\mathrm{T} > 2$~GeV/$c$ for both PbSc and PbGl.

\begin{figure}[tb]
\includegraphics[width=1.0\linewidth]{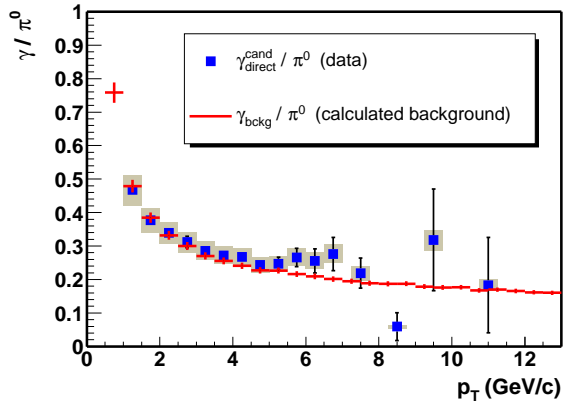}
\caption{\label{fig1} (color online)
Ratio of direct-photon candidates to the $\pi^0$ spectrum. 
The histogram represents the expected background signal from the Monte Carlo 
calculation. The error bars represent the statistical error and the boxes the 
systematic error.}
\end{figure} 

\begin{table}
\caption{\label{tab:syserr}
Systematic uncertainties of the neutral-pion spectrum, 
the direct-photon-candidate spectrum, and the measured and simulated 
$\gamma/\pi^0$ ratios at $p_\mathrm{T} = 6.75$~GeV/$c$.}
\begin{ruledtabular} \begin{tabular}{lcc}
$\pi^0$ error source & \hspace{15pt}PbGl\hspace{15pt} &  \hspace{15pt}PbSc\hspace{15pt} \\
\hline
Yield extraction & 5\,\% & 5\,\%\\
Yield correction & 8\,\% & 6\,\% \\
Energy scale     & 9\,\% & 8\,\% \\
\hline
Total & 13\,\% & 12\,\% \\
\hline\hline
$\gamma_\mathrm{direct}^\mathrm{cand}$ error source &  PbGl &  PbSc \\
\hline
Non-$\gamma$ background correction & 4\,\% & 4\,\% \\
Yield correction & 6\,\% & 5\,\% \\
Energy scale     &  9\,\% & 9\,\% \\
\hline
Total & 12\,\% & 11\,\%\\
\hline\hline
Total error & \multicolumn{2}{c}{PbGl+PbSc combined} \\
\hline
$\gamma_\mathrm{direct}^\mathrm{cand}/\pi^0$ &  \multicolumn{2}{c}{9\,\%}\\
$ \gamma_\mathrm{bckg}/\pi^0$ &  \multicolumn{2}{c}{4\,\%}\\
\end{tabular} \end{ruledtabular} 
\end{table}

The geometric acceptance and the efficiency of the photon detection
were calculated with a Monte Carlo simulation. The efficiency takes
the distortion of the direct-photon-candidate spectrum due to energy
smearing into account. Moreover, it corrects the small ($\lesssim
5\%$) loss of photons due to the shower shape and time-of-flight cuts.
The final direct-photon spectrum was corrected for the difference
between the average direct-photon cross section within a finite
$p_\mathrm{T}$ bin and the value of the cross section at the bin
center.

With the described corrections the unbiased differential cross section
$\gamma_\mathrm{direct}^\mathrm{cand} \equiv E \, \mathrm{d}^3
\sigma/\mathrm{d}^3p$ for the direct-photon candidates calculated from
the minimum bias data sample reads

\begin{equation}
E \frac{\mathrm{d}^3 \sigma}{\mathrm{d}^3p} =
  \frac{1}{\hat{\mathcal{L}}} \cdot 
  \frac{1}{2 \pi p_\mathrm{T}} \cdot
  \frac{C_\mathrm{reco} \cdot C_\mathrm{conv} \cdot C_\mathrm{loss}}{f} \cdot
  \frac{N_{\mathrm{direct}\,\gamma}^\mathrm{cand}}{\Delta p_\mathrm{T} \cdot \Delta y} 
\end{equation}

where $N_{\mathrm{direct}\,\gamma}^\mathrm{cand}$ is the total number
of direct-photon candidates in a $p_\mathrm{T}$ bin $\Delta
p_\mathrm{T}$ and rapidity bin $\Delta y$; $C_\mathrm{reco}$ is the
acceptance and efficiency correction for photons; $C_\mathrm{conv}$ is
the correction for photon conversions; $C_\mathrm{loss}$ is the
correction for the loss of genuine direct photons in the $\pi^0$ and
$\eta$ tagging; $f = 0.75 \pm 0.02$ is the fraction of the unbiased
direct photon yield which is measured under the minimum bias trigger
condition; and $\hat{\mathcal{L}}$ is the integrated luminosity for
the analyzed data sample. The high-$p_\mathrm{T}$ triggered sample
required an additional correction for the efficiency of this trigger
for photon detection.

The $p_\mathrm{T}$ spectrum of the direct-photon candidates contains
direct photons as well as remaining background photons from hadron
decays. These background photons mostly come from $\pi^0$ and $\eta$
decays for which one decay photon misses the detector. At a
representative bin of $p_\mathrm{T} = 6.75$~GeV/$c$ about 93\% of the
background photons originate from $\pi^0$ and $\eta$ decays, the
remaining background photon come from decays of other hadrons like
$\omega$ and $\eta'$. The background was calculated with the same
Monte Carlo code that was used for the acceptance and efficiency
calculation. The Monte Carlo code took a parameterization of the
measured $\pi^0$ spectrum as input. The spectra of $\eta$ mesons and
other hadrons with photon decay branches were assumed to have the same
shape as the $\pi^0$ spectrum as a function of $m_\mathrm{T} =
\sqrt{p_\mathrm{T}^2+m_0^2}$ ($m_\mathrm{T}$ scaling
\cite{Alper:1974rw,Alpgard:1981ke}). The $\eta/\pi^0$ invariant cross
section ratio as a function of $m_\mathrm{T}$ was taken as $0.48 \pm
0.1$ which was confirmed by the measured $\eta$ spectrum.

The dominant systematic uncertainty of the neutral-pion spectrum and
the direct-photon-candidate spectrum came from the uncertainty of the
energy scale and the uncertainty of the yield correction. At
$p_\mathrm{T} = 6.75$~GeV/$c$, the 1.5\% uncertainty of the energy
scale resulted in a 9\% uncertainty in the photon yield. The yield
correction included the correction for energy and position smearing of
the detector, for photon losses due to particle-identification cuts,
for photon conversions, and for the detector acceptance. For the
neutral pions an additional 5\% uncertainty came from the extraction
of the $\pi^0$-peak content. In the case of the photon measurement the
uncertainty due to the subtraction of charged and neutral backgrounds
was taken into account.  In the ratio
$\gamma_\mathrm{direct}^\mathrm{cand}/\pi^0$ of the
direct-photon-candidate spectrum and the neutral-pion spectrum
systematic uncertainties partially cancel. Monte Carlo studies showed
that the uncertainty of this ratio at $p_\mathrm{T} = 6.75$~GeV/$c$
due to a possible non-linearity of the energy scale was $\sim$2\%.

For the determination of the direct-photon spectrum the expected
background photons from hadronic decays need to be subtracted from
the spectrum of direct-photon candidates. To this end the ratio
$R_\gamma = (\gamma_\mathrm{direct}^\mathrm{cand}/\pi^0) /
(\gamma_\mathrm{bckg}/\pi^0)$ of the measured direct-photon candidates
to the calculated background was determined. The direct-photon
spectrum was then calculated as

\begin{equation}
\gamma_\mathrm{direct} = (1 - R_{\gamma}^{-1}) \cdot 
\gamma_\mathrm{direct}^\mathrm{cand}.
\label{eq:gamma}
\end{equation}

\begin{figure}[tb]
\includegraphics[width=1.0\linewidth]{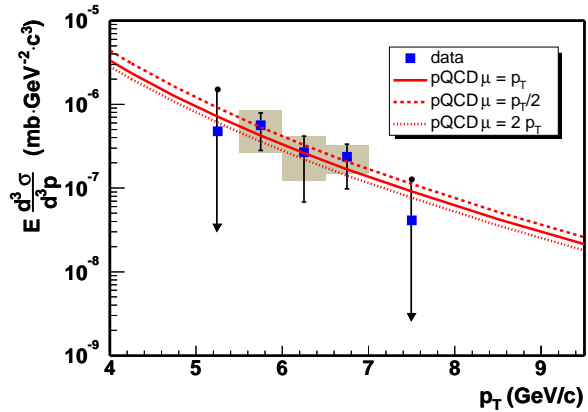}
\caption{\label{fig2} (color online)
  Measured cross section and NLO pQCD 
  calculations for direct-photon production in p+p collisions at
  $\sqrt{s}=$200~GeV. The normalization error of 9.6\% is not shown.
  The two data points plotted with an arrow indicate the beginning of
  the low- and high-$p_\mathrm{T}$ ranges where the direct photon
  signal is consistent with zero. The upper edges of the arrows
  indicate an upper limit (90\% confidence level) for the direct
  photon cross section calculated from the statistical and systematic
  uncertainty. }
\end{figure} 

The relative systematic uncertainty of the direct-photon cross section
was calculated as the quadratic sum of the relative uncertainties of
the two factors in Eq.~\ref{eq:gamma}.  The factor $1-R_{\gamma}^{-1}$
contains the (statistical and systematic) significance of the direct
photon signal. When multiplying with
$\gamma_\mathrm{direct}^\mathrm{cand}$, only the systematic
uncertainties that cancelled in the ratio $R_{\gamma}$ are added ({\em
  e.g.}  the energy scale error). The overall normalization uncertainty
from the luminosity determination was 9.6\%. The estimated systematic
uncertainties for the measured neutral-pion spectrum and the measured
direct-photon-candidate spectrum are shown in Table~\ref{tab:syserr}
for a representative bin ($p_\mathrm{T} = 6.75$~GeV/$c$) of the
spectrum.  The systematic uncertainty of the direct-photon measurement
was corroborated by comparing results obtained for the different
photon-identification criteria. Moreover, the individual PbGl and PbSc
results were found to agree within systematic errors.

To make the best use of the available Run-2 statistics the photon and
$\pi^0$ spectra from PbSc and PbGl were combined for the final result.
The ratio of the acceptance- and efficiency-corrected direct-photon
candidate $p_\mathrm{T}$ spectrum to the measured $\pi^0$ spectrum is
depicted in Fig.~\ref{fig1}. In addition, the Monte Carlo calculation
for the ratio of the expected background photons to the $\pi^0$
spectrum is shown.  At high $p_\mathrm{T}$ the statistical
significance of the direct-photon-candidate spectrum is weak. However,
around $p_\mathrm{T} \approx 6-7$~GeV/$c$ there is clear evidence of a
photon signal above the background.

\begin{table}
\caption{Invariant differential cross section for direct-photon
production in p+p collisions at $\sqrt{s}=200$~GeV. Asymmetric 
uncertainties ($\sigma_\mathrm{low}$, $\sigma_\mathrm{high}$) are
given for the cross section. The absolute
normalization error of 9.6\% is not included.}
\begin{ruledtabular} \begin{tabular}{cccccc}
$p_\mathrm{T}$ & $E \, \mathrm{d}^3\sigma/\mathrm{d}^3p$ & 
    \multicolumn{2}{c}{stat. error} & \multicolumn{2}{c}{sys. error}\\ 
(GeV/$c$) & (mb GeV$^{-2}c^3$) & 
  $\sigma_\mathrm{low}$ & $\sigma_\mathrm{high}$& 
  $\sigma_\mathrm{low}$ & $\sigma_\mathrm{high}$\\
\hline
5.75 & 5.61$\cdot 10^{-7}$& 50\,\% & 42\,\% & 53\,\% & 54\,\%\\
6.25 & 2.68$\cdot 10^{-7}$& 75\,\% & 56\,\% & 55\,\% & 56\,\%\\
6.75 & 2.37$\cdot 10^{-7}$& 59\,\% & 42\,\% & 37\,\% & 38\,\%\\
\end{tabular} \end{ruledtabular} 
\label{tab:result}
\end{table}

The extracted invariant direct-photon spectrum is shown in
Fig.~\ref{fig2} and the numerical values are given in
Table~\ref{tab:result}.  The experimental result was compared to NLO
pQCD calculations \cite{Aurenche:1983ws, Aurenche:1987fs, Baer:1989xj,
  Baer:1990ra, Gordon:1993qc, Gordon:1994ut} which used the CTEQ6
parton distribution functions \cite{cteq6} and the GRV
parton-to-photon fragmentation function \cite{Gluck:1992zx}. There are
in general two mechanisms for the production of direct photons: the
direct contribution from elementary scattering processes of quarks and
gluons, described in the introduction, and the contribution from
photons which are produced in the fragmentation of quark or gluon
jets.  The latter is a long-distance process which is not
perturbatively calculable. It is described by a parton-to-photon
fragmentation function which is determined experimentally. Since no
isolation cut was used in the data analysis the pQCD calculation in
Fig.~\ref{fig2} includes contributions from the direct production
mechanism and the fragmentation mechanism.  The separation of
short-distance and long-distance processes in the pQCD calculation
introduces unphysical renormalization, factorization, and
fragmentation scales. Identical values for all three scales were used
in the pQCD calculation. In Fig.~\ref{fig2}, results are shown for
three choices of the scales ($\mu = p_\mathrm{T}$, $\mu =
p_\mathrm{T}/2$, and $\mu = 2 p_\mathrm{T}$).  The theoretical and
experimental results agree within the large uncertainties of the data
points.

Prior to the Run-2 p+p beamtime PHENIX took data from Au+Au collisions
at $\sqrt{s_\mathrm{NN}}=200$~GeV. A clear direct-photon signal was
observed in mid-central and central Au+Au reactions
\cite{Frantz:2004gg}. The strong suppression of $\pi^0$'s and $\eta$'s
in Au+Au significantly reduced the number of background photons and
eased the extraction of the direct-photon signal.  The p+p NLO pQCD
was used as a baseline reference for the interpretation of the Au+Au
result. In contrast to neutral pions no sign of a suppression of
direct photons in Au+Au collisions was found.  The direct-photon
measurement presented in this paper supports the use of the NLO pQCD
calculation as a reference for the results measured in Au+Au.

In summary, a small but significant direct-photon signal has been
observed at mid-rapidity in p+p collisions at $\sqrt{s}=200$~GeV. The
measured direct-photon cross section is in agreement with pQCD calculation,
albeit within large errors.


We thank the staff of the Collider-Accelerator and Physics
Departments at BNL for their vital contributions.  We acknowledge
support from the Department of Energy and NSF (U.S.A.), 
MEXT and JSPS (Japan), CNPq and FAPESP (Brazil), NSFC (China), 
CNRS-IN2P3 and CEA (France), 
BMBF, DAAD, and AvH (Germany), 
OTKA (Hungary), DAE and DST (India), ISF (Israel), 
KRF and CHEP (Korea), RMIST, RAS, and RMAE (Russia), 
VR and KAW (Sweden), U.S. CRDF for the FSU, 
US-Hungarian NSF-OTKA-MTA, and US-Israel BSF.


\end{document}